\documentclass[conference]{IEEEtran}
\IEEEoverridecommandlockouts

\usepackage{cite}
\usepackage{amsmath,amssymb,amsfonts}
\usepackage{algorithmic}
\usepackage{graphicx}
\usepackage{textcomp}
\usepackage{xcolor}
\usepackage{url}
\usepackage{comment}
\usepackage{booktabs}
\usepackage{etoolbox}
\AtBeginEnvironment{thebibliography}{\tiny}

\usepackage{setspace}
\def\BibTeX{{\rm B\kern-.05em{\sc i\kern-.025em b}\kern-.08em
    T\kern-.1667em\lower.7ex\hbox{E}\kern-.125emX}}
\begin{document}

\title{OASIS: \underline{O}ptimized Lightweight \underline{A}utoencoder \underline{S}ystem for Distributed \underline{I}n-\underline{S}ensor computing\\
\vspace{-2mm}
}

\author{
\IEEEauthorblockN{
Chengwei Zhou$^{1}$,
Sreetama Sarkar$^{2}$,
Yuming Li$^{2}$,
Arnab Sanyal$^{3}$,
Md Abdullah-Al Kaiser$^{4}$, \\
Sercan Aygun$^{5}$, 
Hassan Najafi$^{1}$,
and Gourav Datta$^{1}$
}

\IEEEauthorblockA{
$^{1}$Case Western Reserve University, USA \
$^{2}$University of Southern California, USA \
$^{3}$University of Texas at Austin, USA \\
$^{4}$University of Wisconsin--Madison, USA \
$^{5}$University of Louisiana at Lafayette, USA
}

\IEEEauthorblockA{
\{chengwei.zhou, mhassan.najafi, gourav.datta\}@case.edu,
\{sreetama, liyuming\}@usc.edu, \\
sanyal@utexas.edu,
mkaiser8@wisc.edu,
sercan.aygun@louisiana.edu
}
\vspace{-9mm}
}

\maketitle
\begin{abstract}
In-sensor computing, which integrates computation directly within the sensor, has emerged as a promising paradigm for machine vision applications such as AR/VR and smart home systems. By processing data on-chip before transmission, it alleviates the bandwidth bottleneck caused by high-resolution, high-frame-rate image transmission. We propose a system architecture that integrates a CMOS image sensor (CIS) with a logic chip via advanced packaging, where the logic chip executes a lightweight encoder that aggressively compresses activations before off-chip transmission. However, conventional deep neural network (DNN) partitioning strategies require multiple layers before achieving meaningful dimensionality reduction, limiting bandwidth savings. To address this, we propose a dual-branch autoencoder-based vision architecture trained with a triple loss function combining task-specific, entropy, and reconstruction objectives to produce compact, Huffman-codable representations. The encoder is prototyped on a FPGA to provide hardware-validated energy measurements, complementing circuit-simulated CIS front-end characterization. Our approach achieves a four-order-of-magnitude reduction in transmitted data volume compared to raw input images, resulting in $2{\times}{-}4.5{\times}$ energy savings. We evaluate on CNN and ViT-based models across diverse tasks, achieving state-of-the-art accuracy with energy efficiency of up to 22.7 TOPS/W.
\end{abstract}


\vspace{-2mm}
\section{Introduction}

Computer vision applications are increasingly prevalent in surveillance, disaster management, autonomous driving, and smartphones, driven by advances in image sensing and deep learning. However, vision sensing and processing remain physically separated: CIS converts light into digital pixels \cite{image_sensor}, while computation occurs on remote CPU/GPU platforms, creating bottlenecks in throughput, bandwidth, and energy efficiency. To address these limitations, researchers have explored near-sensor processing \cite{pinkhan2021jetcas,sony2020vision}, in-sensor processing \cite{chen2020pns}, and in-pixel processing \cite{scamp2020eccv}. Near-sensor approaches integrate accelerators on the same board \cite{pinkhan2021jetcas} or through 3D stacking with CIS \cite{sony2020vision}, but still incur significant data-transfer overhead. In-sensor processing \cite{chen2020pns} embeds computation within the CIS periphery, while in-pixel processing \cite{scamp2020eccv, datta2022p2m,datta2022p2mdetrack,p2mdac,p2micassp} performs computation directly inside pixels. However, most prior works target simple datasets such as MNIST or low-resolution CIFAR-10 rather than realistic high-resolution CIS workloads. Although effective for early CNN layers, these approaches struggle to support deeper multi-bit and multi-channel DNN operations under strict hardware constraints. As a result, activation compression remains limited, reducing bandwidth savings, while Amdahl’s law constrains overall system efficiency since most DNN layers still execute off-sensor.

To fully harness the energy efficiency benefits of in-sensor computing, we propose integrating a CIS with a dedicated logic chip, as suggested in prior studies \cite{gomez2022distributed,p2micrc}. This integration leverages micro Through-Silicon Vias ($\mu$TSVs) \cite{gomez2022distributed} as cross-chip interconnects, significantly reducing data transfer energy and increasing bandwidth compared to conventional MIPI interfaces. Utilizing this efficient TSV interface, our strategy involves processing as many DNN layers as possible within the logic chip, while adhering to memory footprint and packaging constraints, and aggressively reducing the dimensionality of activations exiting the sensor. However, prior works adopt existing network architectures \cite{gomez2022distributed} that do not prioritize aggressive compression of spatial and channel dimensions, limiting overall efficiency. This raises a fundamental question: \textit{can we design a DNN that is lightweight enough for a resource-constrained sensor-integrated logic chip, yet capable of compressing activations by orders of magnitude without sacrificing downstream task accuracy?}

We answer this in the affirmative by proposing a \textit{dual-branch autoencoder-based vision architecture} that decouples compression from downstream task inference. A lightweight encoder deployed on the logic chip aggressively compresses spatial and channel dimensions, while the task-specific network operates entirely off-chip. The key insight is that the encoder need not solve the downstream task directly, but instead generate a compact representation from which the off-chip model can recover task-relevant features. To achieve this, the encoder is trained using a \textit{triple loss function}: (1) task-specific loss to preserve discriminative features, (2) entropy loss to encourage compressible latent representations suitable for Huffman coding, and (3) reconstruction loss (MSE) through a decoder branch active only during training. This asymmetric design eliminates decoder overhead during inference, enabling aggressive compression within the strict compute and memory constraints of the sensor-integrated logic chip. Unlike JPEG compression~\cite{wallace1991jpeg}, which removes perceptually insignificant information for human viewing, our method optimizes compression specifically for deep learning inference by preserving task-relevant features. Prior work~\cite{torfason2018towards} performs inference directly on compressed representations without decoding, whereas our approach learns an in-sensor compression strategy optimized for bandwidth and energy efficiency while maintaining spatial and channel structures required for downstream models. Another key limitation in prior in-/near-sensor computing works is the lack of \textit{end-to-end hardware-validated prototypes} for practical vision workloads. Most prior studies rely on analytical energy estimates derived from per-MAC energy and operation counts~\cite{scamp2020eccv,datta2022p2m,angizi2022pisa,kaiser2024voltagecontrolledmagnetictunneljunction}, neglecting control logic, memory banking, pipeline stalls, and interconnect overhead. In contrast, we prototype the complete in-sensor digital pipeline, from encoder inference through quantization and Huffman-compressed bitstream generation, on an FPGA using high-level synthesis. Combined with our circuit-simulated CIS front-end (Monte Carlo analysis in 22nm FDSOI), this constitutes the \textit{first end-to-end hardware-backed in-sensor computing pipeline} validated across multiple practical vision applications, providing both measured energy data and a concrete path toward sensor-integrated deployment.

Experimental results show up to four orders-of-magnitude bandwidth reduction over raw image transmission and two orders-of-magnitude reduction compared to Processing-in-Pixel-in-Memory (P$^2$M)-based DNNs \cite{datta2022p2m}. Compared to direct on-sensor encoder execution, our method reduces encoder energy by $1.15\times$ for ResNet and $13.4\times$ for Swin-Transformer variants. Overall, the proposed approach achieves $2\times{-}4.5\times$ lower total system energy with negligible accuracy loss across image classification, hand tracking, and eye tracking tasks.

\section{Proposed Network Architecture \& Algorithm}
\begin{figure}
    \centering
\includegraphics[width=0.45\textwidth]{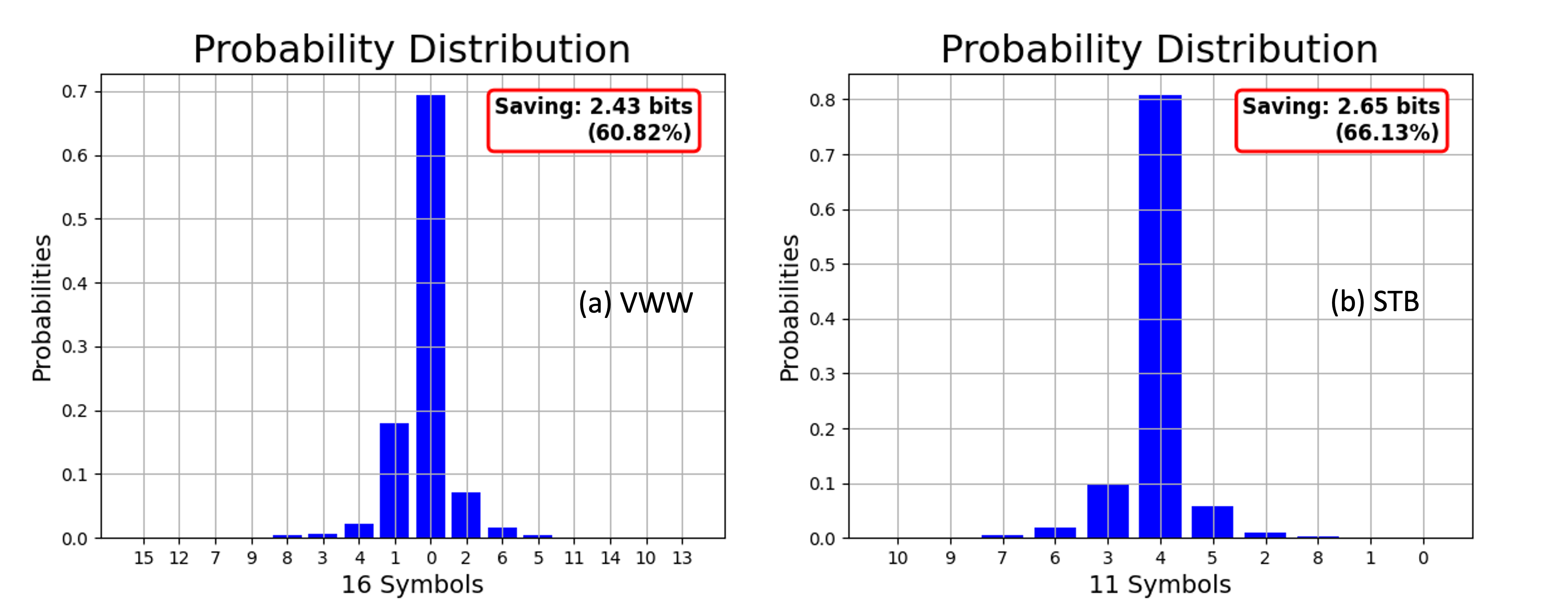}
\vspace{-3mm}
    \caption{\footnotesize Probability distribution of 4-bit encoder output for input images from the evaluation sets of (a) the Visual Wake Words (VWW) dataset, and (b) the Stereo Hand Pose Tracking (STB) dataset.}
    \label{fig:huffman}
    \vspace{-6mm}
\end{figure}

\begin{figure*}
    \centering
\includegraphics[trim={100 235 150 240}, clip, width=\linewidth]{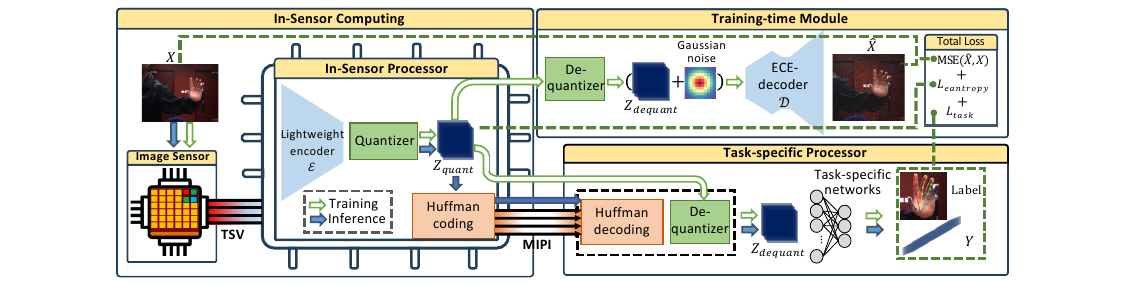}
\vspace{-9mm}
\caption{\footnotesize Overview of the dual-branch autoencoder-based in-sensor computing architecture. The system integrates an \textit{image sensor} with an \textit{in-sensor processor} that implements a lightweight DNN backbone. Quantization and Huffman coding are applied at the processor's output to enhance data efficiency. An off-chip \textit{task-specific processor} handles specialized computations following the encoder. Additionally, a decoder branch is employed exclusively during training for reconstruction supervision, ensuring accurate learning and performance of the autoencoder.
    }
    \label{fig:framework}
    \vspace{-5mm}
\end{figure*}

\noindent\textbf{Dual-Branch Autoencoder-based Network}: We propose a dual-branch autoencoder-based vision network to address challenges in sensor computing. As shown in Fig. \ref{fig:framework}, the framework consists of: (1) a computationally efficient \textit{encoder} \(\mathcal{E}\) on the CIS integrated logic chip, (2) a reconstruction \textit{decoder} \(\mathcal{D}\) activated during training, and de-activated during inference, (3) a task-specific off-chip network for downstream processing. \\
\noindent
\textit{On-chip Encoder Backbone}: The encoder $\mathcal{E}$ generates a low-dimensional latent representation to reduce bandwidth and communication overhead between the sensor and off-chip processors. We develop lightweight ResNet and SwinViT backbones that aggressively compress spatial and channel dimensions while satisfying logic-chip memory constraints and minimizing off-chip DRAM accesses. The ResNet encoder adopts a compact residual design based on \cite{he2016deep}, reducing the stages from four to three and using stride-2 down-sampling. The SwinViT encoder, derived from \cite{liu2021swin}, employs hierarchical down-sampling and increases the patch size from 4 to 8 for stronger compression on high-resolution inputs. To achieve the target compressed size, channel expansion is minimized in early stages and heavily reduced in the final stage, followed by spatial pooling. Finally, the encoder output is quantized and Huffman coded as described in Sections II-B and II-C. \\
\noindent
\textit{Decoder Branch}: We de-quantize the encoder output and inject simulated Gaussian noise to reduce quantization artifacts before passing it to the decoder $\mathcal{D}$ (Fig.~\ref{fig:framework}). The decoder follows an expand-contract-expand (ECE) design that balances computational efficiency and reconstruction quality. The initial expansion captures fine-grained details, the contraction suppresses noise and retains salient features, and the final expansion restores dimensional alignment for reconstruction loss computation. This self-supervised reconstruction preserves task-relevant information for downstream inference.  \\
\noindent
\textit{Off-chip Task-specific Branch}: The off-chip task-specific network performs Huffman decoding and de-quantization to reconstruct the quantized encoder outputs, as shown in Fig. \ref{fig:framework}. It then processes the decoded data based on the specific task, such as a classifier layer for image classification. 

\noindent\textbf{Training Methodology}: Our training method is governed by a joint loss function combining reconstruction, entropy regularization, and task-specific objectives: $\mathcal{L} = \text{MSE}(X, \hat{X}) + \beta\mathcal{L}_{\text{entropy}} + \gamma\mathcal{L}_{\text{task}}$, where $\beta$ and $\gamma$ are hyperparameters. The reconstruction loss minimizes the mean-square error between input images \(X\) and their reconstructed versions \(\hat{X}\), 
The entropy loss is applied to the quantized encoder output to enable efficient compression without significantly compromising performance as follows.
\begin{equation}
\mathcal{L}_{\text{entropy}} = \max(\mathbb{E}_{Z_{\text{quant}} \sim P}[-\log_2 P(Z_{\text{quant}})] - H_{\text{ref}}, 0).
\label{eq:entropy_loss_function}
\end{equation}
Here \(P\) is the quantized output distribution, \(P(Z_{\text{quant}})\) is the probability for the quantized encoder output denoted as \(Z_{\text{quant}}\), and \(H_{\text{ref}}\) controls the rate-distortion trade-off. Following \cite{agustsson2017soft}, we construct the data distribution via histogram analysis. This approach encourages the encoder to produce outputs with lower entropy, making them more amenable to Huffman coding. The task-specific loss $\mathcal{L}_{\text{task}}$ ensures encoder feature learning aligns with the downstream tasks.

We use quantization-aware training (QAT) \cite{jacob2018quantization} that simulates quantization effects during training, enabling the encoder to handle quantization artifacts, and avoid performance drop for the downstream tasks. To address the non-differentiability of quantization, we use the straight-through estimator (STE) \cite{bengio2013estimating}, which can be denoted as $\frac{\partial \mathcal{L}}{\partial Z} = \frac{\partial \mathcal{L}}{\partial Z_{\text{quant}}}$, where $Z$ and $Z_{\text{quant}}$ indicate the full-precision and quantized output of the encoder respectively. We apply symmetric uniform quantization using a momentum-based update strategy during training to dynamically adjust the quantization range. During training, the quantized encoder output is de-quantized back to full-precision value (see Fig. 2) as $Z_{\text{dequant}}{=}Z_{\text{quant}}{\cdot}{q_{\text{scale}}}$, where $q_{\text{scale}}=(\frac{2\cdot{max}}{2^{n}-1})$ denotes the quantization scale. Here, \textit{max} refers to the maximum value of the encoder output's dynamically adjusted range, and $n$ is the bit-precision. 

\noindent\textbf{Huffman Coding}: Huffman encoding, a lossless data compression technique, assigns shorter binary codes to more frequent symbols by constructing a binary tree based on symbol frequencies \cite{huffenc}. In our encoder's quantized outputs, the bell-shaped probability distribution is skewed towards a few values, allowing these frequent symbols to be stored with fewer bits, while rarer symbols require more bits, as illustrated in Fig. \ref{fig:huffman}. This approach reduces data transmission costs while preserving information integrity. Moreover, Huffman encoding guarantees optimality among prefix codes, ensuring no alternative prefix coding scheme can achieve a more compact representation for the given probability distribution.
\vspace{-.5mm}



\section{Proposed Hardware Architecture}
\label{sec:system_arch}

Our in-sensor computing architecture, as illustrated in Fig.~\ref{fig:framework}, integrates an active pixel sensor (APS), an in-sensor processor, and a task-specific processor. We prototype the in-sensor processor on an FPGA to validate feasibility and obtain hardware-measured energy characterization.

\noindent\textbf{Active Pixel Sensor}: The APS array generates analog voltage signals at each pixel, processed in a column-parallel manner with dedicated ADC circuits at each column's end. These signals are converted to digital form and transmitted to the in-sensor processor via TSVs. This approach eliminates the need for a traditional Image Signal Processor (ISP), reducing computational overhead and bandwidth requirements. Moreover, it alleviates the burden on the ADC by operating at a lower bit precision, thereby enhancing energy efficiency without compromising essential feature extraction. Low-power ADC designs, such as successive approximation register (SAR) ADCs, can further optimize resolution and energy efficiency.

\noindent\textbf{Efficient In-Sensor Processor}: The in-sensor processor follows a weight-stationary dataflow strategy to minimize data movement overhead and features a processing-element (PE) based array, where each PE is a quantized multiply-accumulate (MAC) unit with dedicated on-chip memory for storing activations and model weights. Since our encoder backbone is lightweight, all weights fit on-chip, eliminating external DRAM accesses. Tile-based spatial processing dynamically maps activation tiles to the PE array, ensuring high spatial locality, reducing redundant data transfers, and maximizing memory utilization.

We implement the architecture on a Xilinx Zynq UltraScale+ FPGA using Vitis HLS, with on-chip BRAM as weight-stationary memory and pipelined convolutional and linear layers achieving an initiation interval of 1. The encoder output passes through a single-cycle 4-bit quantizer (1 DSP slice) followed by a fully partitioned 16-entry Huffman lookup table that maps symbols to variable-length codewords in one clock cycle; a 64-bit shift register packs the codes into 32-bit DMA words. All stages execute concurrently using HLS \texttt{DATAFLOW} FIFOs, while quantization and Huffman coding incur negligible overhead (${\sim}$250--300 LUTs, 1 DSP). The complete encoder occupies 19,200 LUTs, 48 BRAM18K, and 28 DSP48E2 slices. This FPGA prototype, featuring on-chip weight storage, streaming execution, and DRAM-free operation, provides the measured energy results reported in Section IV. The compressed bitstream is transmitted off-chip via MIPI for task-specific processing.

\vspace{-1mm}
\section{Hardware-Validated System Energy Modeling}
\label{sec:system_modeling}

Accurate energy modeling is critical for evaluating the efficacy of distributed in-sensor computing architectures. Unlike prior works that rely solely on analytical per-operation energy estimates~\cite{gomez2022distributed,datta2022p2m,p2mglsvlsi}, our approach combines \textit{circuit-level simulations} for the analog front-end with \textit{FPGA-based hardware measurements} for the digital compute pipeline, providing a comprehensive and empirically grounded energy characterization of the full system.

\noindent\textbf{Energy Model Formulation}: The total energy per frame, $E_{Total}$, is expressed as $E_{Total}= E_{Aps} + E_{Tsv} + E_{Inf} + E_{Enc} + E_{Back}$, where $E_{Aps}$, $E_{Tsv}$, $E_{Inf}$, $E_{Enc}$, and $E_{Back}$ denote the energy contributions of the APS, TSV interface, sensor-to-processor communication interface, encoder (including quantization and Huffman coding), and the task-specific back-end processor, respectively. The baseline non-in-sensor hardware does not incur $E_{Tsv}$; however, it incurs significantly higher $E_{Inf}$ that dominates $E_{Total}$. 

\noindent\textbf{APS Energy Consumption}: The APS consumes per-frame energy in pixel readout and analog-to-digital conversion: $E_{Aps} = \left({E_{read/pix} + E_{ADC/pix}}\right)\cdot{N_{pix}}$, where $E_{read/pix}$ and $E_{ADC/pix}$ are the per-pixel readout and ADC conversion energies, respectively. These values are derived from our \textit{in-house circuit simulations} of a column-parallel APS in 22nm GlobalFoundries FDSOI technology, incorporating Monte Carlo analysis across process and mismatch variations to capture realistic statistical behavior. From these simulations, $E_{read/pix} + E_{ADC/pix} \approx 63.6$~pJ for 8-bit pixel resolution.

\noindent\textbf{Communication Energy}: Data transfer between hardware components occurs through multiple communication interfaces, each with distinct energy characteristics.

\noindent\textit{1) TSV Energy}: TSVs provide high-bandwidth, low-energy interconnects between the sensor die and the co-packaged logic chip~\cite{Vivet2020}. The TSV communication energy for the input image is $E_{TSV}{=}A_{TSV}{\cdot}E_{Byte,TSV}$, where $A_{TSV}$ is the transmitted data size in bytes and $E_{Byte,TSV}{=}6.25$~pJ/byte~\cite{Vivet2020}. For a $224{\times}224$ image input at 8-bit precision, this yields $E_{TSV} = 0.94~\mu$J.

\noindent\textit{2) MIPI and External Interface Energy}:
The Mobile Industry Processor Interface (MIPI) transfers processed data from the sensor to the off-chip task-specific processor~\cite{Choi2021MIPI}. The energy is given by $E_{Inf}=A_{Size} \cdot E_{Byte,inf}$, where $E_{Byte,inf}{=}100$~pJ/byte~\cite{Choi2021MIPI}. Our autoencoder system dramatically reduces $A_{Inf}$ through aggressive activation compression, directly translating to proportional MIPI energy savings as detailed in Section~\ref{sec:results}. For remote processing scenarios employing wired (Ethernet, USB) or wireless (Wi-Fi, 5G) links, $E_{Byte,inf}$ is substantially higher, further amplifying the energy benefits of our approach.

\noindent\textbf{FPGA-Based Encoder Energy Characterization}: To obtain an accurate and empirically grounded estimate of encoder energy, we measure the power consumption of the complete inference pipeline, including the encoder backbone, quantizer, and Huffman encoder, directly on the FPGA prototype described in Section~\ref{sec:system_arch}. This captures the full system behavior including control logic overhead, memory access patterns, pipeline stalls, and interconnect losses, which are difficult to account for in purely analytical models that estimate compute energy from per-MAC operation counts alone~\cite{datta2022p2m, angizi2022pisa}.

\noindent\textit{1) Power Measurement Methodology}:
Power consumption is estimated using the post-implementation power analysis reports generated by Xilinx Vivado. Dynamic power is obtained from switching-activity-based estimates of the programmable logic (PL) design, while static power corresponds to the device leakage reported by the tool. The reported power values are derived from the final placed-and-routed design operating at the target clock frequency and are used to compare the relative energy efficiency of different encoder configurations. The frame-level encoder energy is computed as $
E_{Enc}^{FPGA} = (P_{active} - P_{idle}) \times T_{frame}$, where $P_{active}$ and $P_{idle}$ denote the measured active and idle PL power respectively, and $T_{frame}$ is the per-frame inference latency.
We estimate the energy consumed by the task-specific back-end ($E_{Back}$) similarly using FPGA measurements.

\noindent\textit{3) Huffman Coding/Decoding Energy}: The Huffman encoder is integrated within the FPGA pipeline and its energy is captured as part of $E_{Enc}^{FPGA}$. For the off-chip Huffman decoder, we estimate the decoding energy based on prior low-power entropy coding implementations \cite{bayar2018low}: $E_{huff{-}dec} \approx 1.15$~pJ/byte. Given the highly compressed bitstream produced by our entropy-regularized encoder (effective bit-width of $\sim$1.57 bits per 4-bit symbol), the Huffman coding overhead is negligible compared to MIPI and compute energy.

\section{Experimental Results}\label{sec:results}

\noindent\textbf{Task Performance}
We evaluate our dual-branch approach on resource-constrained tasks pertinent to smart home and AR applications, specifically focusing on Visual Wake Words (VWW)\cite{chowdhery2019visual} classification, hand tracking, and eye tracking. We quantize the encoder output to 4 bits and train the entire network end-to-end for 200 epochs using AdamW with a cosine learning rate scheduler, setting $H_{ref}=0.7$. For VWW, we set $\beta{=}2$, $\gamma{=}2$, $\text{lr}{=}0.002$ and weight decay of 1e-5, where \textit{lr} denotes the learning rate. For hand tracking, we set $\beta{=}2$, $\gamma{=}4$, $\text{lr}{=}0.1$ and weight decay of 1e-5. For eye-tracking, we set $\beta{=}2$, $\gamma{=}4$, and train our network
using an Adam optimizer with an initial $\text{lr}$ of 1e-3. 

\subsubsection{VWW Classification}
\begin{figure}
    \centering
\includegraphics[width=1\linewidth]{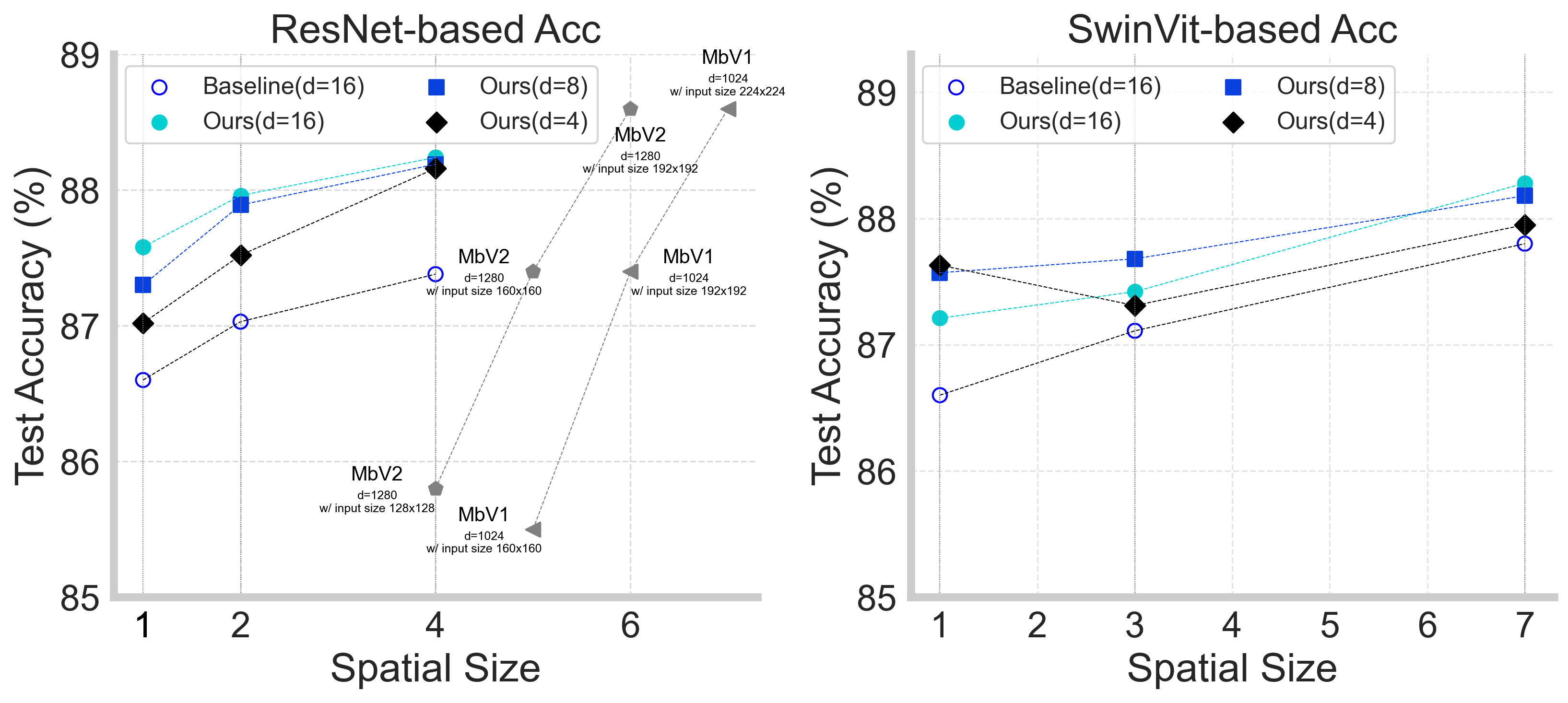}
    \vspace{-7mm}    \caption{Accuracy vs. Representation Dimension comparison on VWW classification task for (\textbf{Left}) ResNet and (\textbf{Right}) SwinViT based architecture, where $d$ denotes the channel dimension of the encoder output.}
    \label{fig:vww}
    \vspace{-4mm}
\end{figure}

VWW focuses on detecting human presence in images, serving as a practical benchmark for lightweight DNNs. We developed compact versions of ResNet and Swin Transformer (SwinViT) architectures for VWW classification, resizing input images to \(224{\times}224\) pixels to aim for high accuracy. \\
\noindent
\textit{a) ResNet-based Autoencoder}: Our compact ResNet-based encoder consists of three layers, each with two residual blocks. The initial convolutional module uses a \(7{\times}7\) kernel with a stride of 4 for aggressive downsampling. The channel dimensions of the three layers are configured as \([128,256,d]\), where \(d\) represents the encoder output channel dimension, which we vary to explore compression benefits. We apply uniform spatial downsampling with factors of 2 per layer, achieving a spatial dimension of \(s{\times}s\) at the encoder output. 
The ECE-decoder progressively reconstructs the feature maps, with layer dimensions \([d,256,128,256,128,64]\) and corresponding spatial scaling factors \([2{\textcolor{green}\uparrow}, 1, 2{\textcolor{green}\uparrow}, 2{\textcolor{red}\downarrow}, 2{\textcolor{green}\uparrow}, 2{\textcolor{green}\uparrow}]\), where \(2{\textcolor{green}\uparrow}\) and \(2{\textcolor{red}\downarrow}\) denote up-sampling and down-sampling by a factor of 2, respectively. As shown in Fig.~\ref{fig:vww}, even with substantial compression, e.g., \(d{=}16\) and \(s{=}1\), our method achieves significantly enhanced performance compared to the baseline configuration, which uses an encoder-only network under the same settings.

\noindent
\textit{b) SwinViT-based Autoencoder}: Our SwinViT-based encoder divides the input image into patches of size \(8{\times}8\) and uses a window size of 7. The encoder comprises four stages (with two blocks per stage) and produces feature maps with dimensions \([64,128,160,d]\) and head number $[2,4,5,4]$ for the four stages. The output of ECE-decoder for each stage is with channel dimensions \([d,160,128,160,128,64]\), head number $[4,5,4,5,4,2]$ and spatial scaling factors \([2\textcolor{green}\uparrow,1,2\textcolor{green}\uparrow,2\textcolor{red}\downarrow,2\textcolor{green}\uparrow,2\textcolor{green}\uparrow]\). Fig.~\ref{fig:vww} shows that our method consistently outperforms the baseline, achieving significant performance gains while maintaining aggressive compression, for different configurations.
\subsubsection{Hand Tracking}
\begin{figure}
    \centering
    \includegraphics[width=1\linewidth]{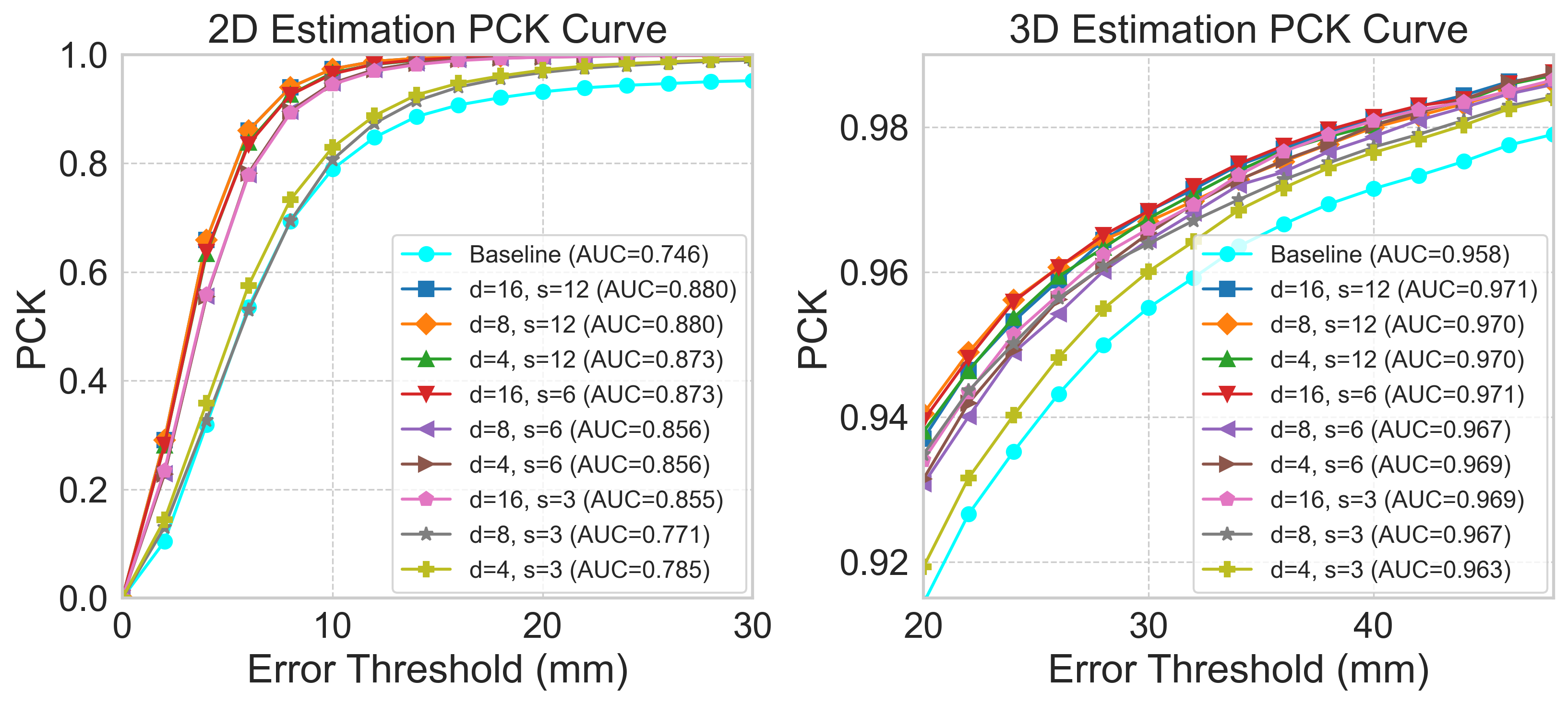}
    \vspace{-7mm}
    \caption{\textbf{Left:} Comparison of 2D estimation AUC and PCK curves, truncated at 30mm, against the baseline.
    \textbf{Right:} Comparison of 3D estimation AUC and PCK curves, truncated within a deviation range of 20mm to 50mm. Baseline model is an encoder-only network with d$=$16, s$=$12.}
    \label{fig:ht}
    \vspace{-5mm}
\end{figure}

We use the Stereo Hand Pose Tracking dataset (STB) \cite{zhang2017hand}, 
which utilizes RGB data from the Intel RealSense F200 camera to estimate both 2D and 3D hand keypoints. The images were center-cropped and resized to \(96{\times}96\) pixels, removing redundant background while preserving the primary hand region.
We adopt the KeyNet-F\cite{han2020megatrack} architecture, modifying the encoder to include four layers with channel dimensions \([32,32,32,64,d]\), where \(d\) represents the encoder output channel dimension, which we vary along with the spatial dimension $s{\times}s$ to explore compression benefits. 

We also develop an additional branch for the ECE-decoder with channel configurations \([d,32,64,32,32]\) and spatial scaling factors \([2{\textcolor{green}\uparrow},2{\textcolor{green}\uparrow},2{\textcolor{red}\downarrow},2{\textcolor{green}\uparrow},1]\). 
The task-specific network comprises a fused network, a keypoint heatmap network for generating 2D heatmaps, and a distance network, implemented to produce distance heatmaps for 3D keypoint estimation. 
Figure~\ref{fig:ht} illustrates the impact of \(d\) and \(s\) on 2D and 3D hand keypoint estimation tasks. For 2D tasks, a higher channel count (\(d{=}16\)) and higher spatial resolution (\(s{=}12\)) yielded an Area Under the Curve (AUC) of 0.880, outperforming the baseline of 0.746. However, reducing the dimension count to \(d{=}4\) and \(s{=}3\) resulted in a lower AUC of 0.785. On 3D tasks, our networks achieved AUC values exceeding the baseline of 0.958. Even under aggressive compression ($d{=}16$ and $s{=}3$), the AUC remained high at 0.969.

\subsubsection{Eye Tracking}
Our eye tracking pipeline consists of two stages: (1) feature extraction for eye segmentation, followed by (2) gaze estimation. We use the EyeNet model as baseline \cite{feng2022edgaze}, a lightweight U-Net based segmentation model, and evaluate our approach on the OpenEDS \cite{palmero2020openeds2020} dataset in Table \ref{tab:eye_tracking}. To reduce data transmission bandwidth, we apply channel reduction from 32 to 4 at the encoder output of EyeNet using our proposed losses. Combined with 4-bit quantization, this achieves a 192$\times$ reduction in the encoder activation dimension as compared to 8-bit image inputs, and 24$\times$ reduction over encoder output size for baseline EyeNet with a mIoU degradation of only 1.4\%, as shown in Table \ref{tab:eye_tracking}. Further compression is possible through region-of-interest (ROI) prediction \cite{kaiser2024energy,feng2022edgaze}, achieving up to $364\times$ overall reduction ($1.9\times$ reduction in input size via ROI). Since ROI prediction relies on previous frames, mIoU may improve over full sequences. However, due to limited annotated frames, evaluation is constrained, leading to mIoU degradation with ROI.

\begin{table}[t!]
    \centering
    \caption{Comparison of Eye segmentation results between EyeNet and our network on OpenEDS dataset w/ and w/o ROI prediction.}
    \label{tab:eye_tracking}
    \begin{tabular}{|l|c|c|c|}
    \hline
        \textbf{Model} & \textbf{Inp. Dim.} & \textbf{Enc. Out Dim.} & \textbf{mIoU} \\ \hline
        EyeNet & (1, 400, 640) & (32, 25, 40) & 0.988\\ \hline
        EyeNet (w/ ROI) & (1, 240, 560) & (32, 15, 35) & 0.967\\ \hline
        Ours & (1, 400, 640) & (4, 25, 40) & 0.974\\ \hline
        Ours (w/ ROI) & (1, 240, 560) & (4, 15, 35) & 0.939\\ \hline
    \end{tabular}
    \vspace{1mm}
    \vspace{-5mm}
\end{table}

\begin{figure}
\centering
\includegraphics[width=0.98\linewidth]{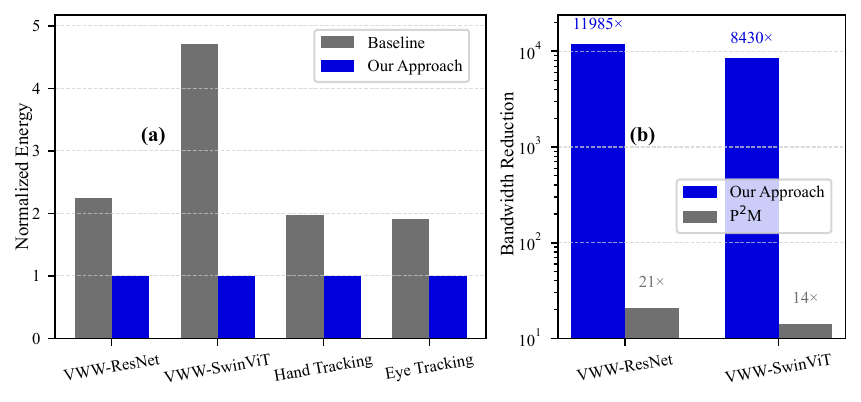}
\vspace{-5mm}
\caption{(a) Normalized energy consumption comparison between our approach and baseline across vision tasks. (b) Bandwidth reduction relative to input image for our approach and P$^2$M (reported only for VWW, the common benchmark in \cite{datta2022p2m}). Identical first-layer strides are used in ResNet \& SwinViT.}
\label{fig:energy_comparison}
\vspace{-6mm}
\end{figure}

\noindent\textbf{Energy Efficiency}:
We compute the total energy consumed by our in-sensor computing hardware and baseline hardware using the energy model developed in Section IV. The energy components differ in several important aspects. The APS energy in Section IV-B remains identical for both hardware implementations. The in-sensor computing hardware incurs an additional TSV energy of $0.94\,\mu\text{J}$ for an input resolution of $224{\times}224$ for internal data transmission between the sensor and logic chip, which the baseline hardware does not incur. However, our in-sensor computing hardware achieves significantly lower MIPI interface energy due to the substantial activation dimension compression enabled by our autoencoder system. For instance, in the VWW task, our encoder with spatial size $4{\times}4$ and 4 channels achieves a dimension reduction factor of $\frac{224{\times}224{\times}3}{4{\times}4{\times}4}{=}2352$. Combined with 4-bit quantization (versus 8-bit input) and Huffman coding (effective bit-width of 1.57 bits), the total bandwidth reduction is $2352{\times}\frac{8}{4}{\times}\frac{4}{1.57}{\approx}11985{\times}$. This reduces the dominant MIPI interface energy from 99.5$\mu$J to a negligible value, yielding substantial total energy savings. Similar trends hold across other tasks and models.

As shown in Fig.~\ref{fig:energy_comparison}(a), our approach reduces the vision pipeline energy by ${\sim}2\times$ compared to baseline non-in-sensor systems across most tasks, with an exceptional ${\sim}4.5\times$ reduction for the SwinViT-based VWW task. This greater efficiency stems from our tiny SwinViT encoder, which significantly reduces the multi-head self-attention module size, drastically lowering the encoder energy by $13.4\times$ (our ResNet encoder reduces energy only by $1.15\times)$ and peak memory requirements. Unlike traditional DNN backbones that struggle to aggressively reduce activation dimensions and often require external DRAM access due to their larger memory footprint, our autoencoder-based approach optimizes the vision pipeline while respecting the strict compute and memory constraints of edge devices. Moreover, as shown in Table \ref{tab:comparison}, our approach achieves 22.7 TOPS/W, nearly twice the efficiency of prior works (11.49 TOPS/W in DPCE \cite{dpce}), while maintaining high accuracy. Unlike DPCE's binary multiplications, our method supports full multi-bit operations for complex convolution and self-attention functions. Compared to analog in-sensor computing systems (e.g., P$^2$M \cite{datta2022p2m}) that implements only the first convolutional block within the sensor, our method achieves a $570.7\times$ ($\frac{11985}{21}\times$ as seen in Fig. ~\ref{fig:energy_comparison}(b)) bandwidth reduction with the same input resolution, leading to substantial energy and latency savings. Moreover, unlike such analog implementations that have limited reconfigurability, our digital domain approach enables greater flexibility and adaptation across various vision tasks. Additionally, while \cite{chen2020pns} works with low-resolution datasets (MNIST, CIFAR-10), our method effectively processes high-resolution ($224{\times}224$) images critical for real-world applications, such as AR.

\begin{table}[t]
\centering
\caption{Comparison of in-sensor computing approaches with varying DNNs, tasks, and CIS technology nodes.}
\vspace{-2mm}
\label{tab:comparison}
\footnotesize
\setlength{\tabcolsep}{1.8pt}
\begin{tabular}{|l|c|c|c|c|c|c|}
\hline
\textbf{Method} & \textbf{Task} & \textbf{Res.} & \textbf{Tech.} & \textbf{Network} & \textbf{TOPS/W} & \textbf{Acc.} \\ \hline
Senputing \cite{Xu2022} & MNIST & 28² & 180nm & 2-layer MLP & 4.7 & 93.76\% \\ \hline
SCAMP \cite{scamp2020eccv} & MNIST & 256² & 180nm & 2-layer CNN & 0.535 & 93.0\% \\ \hline
DPCE \cite{dpce} & CIFAR10 & 32² & - & LeNet-5 & 11.49 & 87.20\% \\ \hline
PIPSIM \cite{pipsim} & CIFAR10 & 64² & 45nm & LeNet-5 & 4.12 & 90.05\% \\ \hline
P$^2$M \cite{datta2022p2m} & VWW & 224² & 22nm & MobileNetV2 & 0.4 & 84.3\% \\ \hline
\textbf{Ours} & VWW & 224² & 22nm & {Tiny SwinViT} & \textbf{15.5} & {88.2\%} \\ \hline
\textbf{Ours} & Hand Track & 96² & 22nm & {KeyNet-F} & \textbf{22.7} & {96.9\%} \\ \hline
\multicolumn{7}{l}{$^*$\scriptsize{1 OP = 1 MAC between weight and input activation in a DNN.}}
\end{tabular}
\vspace{-7mm}
\end{table}

\noindent\textbf{Latency}:
Our bandwidth reduction also directly translates to latency savings, as MIPI transmission dominates end-to-end latency in conventional pipelines. The TSV interface provides 200$\times$ greater bandwidth than MIPI~\cite{Choi2021MIPI}, making sensor-to-logic-chip transfer negligible compared to baseline MIPI.

\section{Conclusions \& Discussions}

We presented OASIS, an optimized lightweight autoencoder for distributed in-sensor computing that significantly improves bandwidth efficiency, energy consumption, and latency for edge AI applications. By combining an asymmetric autoencoder architecture with entropy-regularized quantization and FPGA-validated energy modeling, our approach achieves up to 11,985$\times$ bandwidth reduction and 2$\times$--4.5$\times$ energy savings over non-in-sensor baselines, with energy efficiency of 22.7 TOPS/W. Future directions include adaptive task-aware compression that adjusts encoder bitrate to scene complexity, and integration of emerging non-volatile memories (e.g., ReRAM) to further reduce energy consumption. 

{\footnotesize 
\bibliographystyle{unsrt}
\bibliography{references_ISVLSI}
}

\end{document}